\documentclass[twocolumn,showpacs,showkeys,floatfix,amsmath,amssymb,prb]{revtex4}% Physical Review B

\usepackage{graphicx}% Include figure files
\usepackage{dcolumn}% Align table columns on decimal point
\usepackage{bm}% bold math
\usepackage{amsmath}

\newcommand{\ro}{\bm{\rho}}

\newcommand{\bPi}{\bm{\Pi}}
\newcommand{\bn}{\bm{\nabla}}
\begin{document}

%\preprint{\large\it  Prepared for submittal to Phys. Rev. B}
%\wideabs{

\title{Interacting like charges in Landau levels: Planar geometry, \\  
symmetries, and  effective quasiparticles }
 \author{A. B. Dzyubenko}
\affiliation{Department of Physics, California State University at Bakersfield, 
Bakersfield, CA 93311, USA} 
\email{adzyubenko@csub.edu}
\altaffiliation[]{on leave from General Physics Institute, RAS,
               Moscow 117942, Russia.}

\author{A. R. Todd}
\affiliation{Department of Physics, California State University at Bakersfield, 
Bakersfield, CA 93311, USA}

\date{\today}
\begin{abstract}
We consider a system of two interacting particles with like but unequal charges
in a magnetic field in the planar geometry. We construct a complete basis of states 
compatible with both the axial symmetry and magnetic translations. The basis is obtained
using a canonical transformation that generates effective quasiparticles
with modified interactions. We establish a connection of this transformation
with the $SU(2)$ algebra and make use of the $SU(2)$ Baker-Campbell-Hausdorff formulas
for evaluating the interaction matrix elements. 
We calculate analytically the eigenenergies of the problem (Haldane pseudopotentials) 
in the first few Landau levels 
for a relatively wide class of interaction potentials. 
\end{abstract}



\maketitle

%%%%%%%%%%%%%%%%%%%%%%%%%%%%%%%%%
\section{Introduction}

Interparticle correlations are enhanced in the presence of a strong magnetic field and 
play a central role in determining the physics of various systems,
such as, atoms and ions in ultrastrong magnetic fields\cite{Rud94,Sim78,Hir83}
and electrons in Landau levels.\cite{QHE1,QHE}
The corresponding few- and many-body electron problems have been considered in detail
in both planar  and spherical  geometries, 
especially in the context of the fractional quantum 
Hall effect.\cite{Lau83,Lau83b,Hal83,Hal87,QHE1,QHE}

In this work, we consider a system of two like but unequal charges in a magnetic field 
interacting via an arbitrary pairwise potential. Our motivation is symmetry-related: 
to the best of our knowledge, this problem has not been treated 
with a full account of the relevant symmetries in the planar geometry.  
At the same time, such systems present physical interest and may 
describe, e.g., multicharged complexes in layered systems\cite{Yud96}  and 
strongly bound ions interacting with one another and being treated as 
structurless point-like charges.\cite{Hir83} 
The developed algebraic approach may also be useful for studying interactions 
between fractionally charged quasiparticles\cite{QHE,Jain05,Sim07}  
and bound states consisting of such excitations.\cite{Ras93,Par03,Wojs06}

This approach may also be relevant for separating 
variables in bulk three-dimensional systems\cite{Sim78,Hir83} and in 
electron systems consisting of two groups of electrons with, e.g., different masses.\cite{Asa98} 
In the latter case the charges and coordinates 
introduced in Sec.~\ref{subsec:Ham} below correspond to the total charges and 
center-of-charge coordinates of the two subsystems.\cite{Dzy00}

%%%%%%%%%%%%%%%%%%%%%%%%%%%%%%%%%
\section{Two interacting charged particles in a magnetic field}

\subsection{The Hamiltonian and Bose ladder operators}
  \label{subsec:Ham}

We consider a system ot two interacting like charges moving on a 2D plane
perpendicular to a uniform magnetic field ${\bf B}=(0,0,B)$.
The latter is assumed to be strong so that interaction-induced
mixing of Landau levels (LLs) is weak and, in the first approximation, can be ignored.  
Our goal here is to construct a complete basis of states that is compatible
with all symmetries of the problem and to find exact diagonalization 
for interacting states in a few lowest LLs. Note that in the absence of the 
interparticle interaction, the states are infinitely-fold degenerate.
Our procedure allows one to consider mixing between LLs perturbatively.
 
To be definite, let us take the charges to be negative, 
$-q_1 <0$ and $-q_2<0$. \cite{Note:Charge}
The Hamiltonian describing the system is
\begin{equation}
                \label{calH} %(2)
  \mathcal{H} = \frac{\hat{\bPi}_{1}^2}{2m_1} + \frac{\hat{\bPi}_{2}^2}{2m_2} 
       + V(| {\bf r}_1 - {\bf r}_2 |)  
\equiv \mathcal{H}_{01} + \mathcal{H}_{02} + V_{\rm int} \; .
\end{equation}
Here $\hat{\bPi}_j = -i\hbar \bn_j +  \frac{q_j}{c} {\bf A}({\bf r}_j)$
are kinematic momentum operators and
$V(r)$ is the interaction potential that can be rather arbitrary.
In the symmetric gauge
${\bf A} = \frac12 {\bf B} \times {\bf r}$
the states of free non-interacting particles in LLs can be described 
by the factored wave functions\cite{QHE,QHE1,Dzy00} 
\begin{equation}
              \label{e_LL}
\phi_{n m}^{(j)}({\bf r}_j)=\frac{1}{\sqrt{n!m!}}
\langle {\bf r}_j |(A^{\dag}_j)^n (B^{\dag}_j)^m |0\rangle \; ,
\end{equation}
constructed using the raising Bose ladder inter- and intra-LL operators,
correspondingly,
\begin{eqnarray}
        \label{lad_en}
    A^{\dag}({\bf r}_j)   & = & \frac{1}{\sqrt{2}}
  \left( \frac{z_j}{2l_j} -
       2l_j \frac{\partial \hphantom{,}} {\partial z_j^{\ast}} \right) \; , \\
      \label{lad_em}
    B^{\dag}({\bf r}_j)  & = & \frac{1}{\sqrt{2}}
  \left( \frac{z_j^{\ast} }{2l_j} -
       2l_j \frac{\partial \hphantom{,} }{\partial z_j} \right) \; .
\end{eqnarray}
Here $ z_j = x_j + iy_j$ are the 2D complex coordinates and
$l_j = (\hbar c/q_j B)^{1/2}$ are the magnetic lengths.\cite{Note1}
The operators commute as $[A_j,A_k^{\dag}]=\delta_{j,k}$, $[B_j,B_k^{\dag}]=\delta_{j,k}$,
and $[A_j,B_k^{\dag}]=[A_j,B_k]=0$. 
The inter-LL operators are connected with the kinematic momentum operators as  
$A^{\dag}_j = 
 - i  \, (\hat{ \Pi }_{jx} + i \hat{ \Pi }_{jy}) l_j/\sqrt{2}\,\hbar$.
The symmetry and algebra leading to the intra-LL operators $B^{\dag}_j$ is discussed below.  

The Hamiltonian of free particles from (\ref{calH}) 
expressed in terms of the inter-LL operators becomes
\begin{eqnarray}
        \label{Ham_0}
\mathcal{H}_{0} \equiv \mathcal{H}_{01} + \mathcal{H}_{02} & = & 
     \hbar\omega_{1} \left( 
            A^{\dag}({\bf r}_1) A({\bf r}_1) + \tfrac{1}{2} \right)  \\
    \nonumber
  & &  + {\mbox \,\,\, }
            \hbar\omega_{2} \left( 
            A^{\dag}({\bf r}_2) A({\bf r}_2) + \tfrac{1}{2} \right)  \; ,
\end{eqnarray}
where $\omega_j = q_j B/ m_j c$ are the cyclotron frequencies.
A complete basis of states in LLs can be constructed as 
\begin{eqnarray}
        \label{basis}
 & \mbox{} & |n_1, n_2, m_1, m_2 \rangle =  
\frac{1}{ \left( n_1! \, n_2! \, m_1! \, m_2! \right)^{1/2} } \times \\
             \nonumber
      & \times &  \left[ A^{\dag}({\bf r}_1) \right]^{n_1} 
                  \left[ A^{\dag}({\bf r}_2) \right]^{n_2} 
                  \left[ B^{\dag}({\bf r}_1) \right]^{m_1} 
                  \left[ B^{\dag}({\bf r}_2) \right]^{m_2} | 0 \rangle \; .
\end{eqnarray}
Here $| 0 \rangle$ is a short-hand notation for the 
vacuum state $| 0, 0, 0,0 \rangle$ for all annihilating operators.   
 
%%%%%%%%%%%%%%%%%%%%%%%%%%%%%%%%%%%%%%%%%%%%%%%%%%%%%%%%%%%%%%%%%%%%%%%%%%%%%%%%%%%%%%%%%
\subsection{Geometrical symmetries}

The system posesses the axial symmetry, $[\mathcal{H},\hat{L}_z]=0$,
where $\hat{L}_z = \sum_j \hat{L}_{\mathrm{zj}} = \sum_j [{\bf r}_j 
\! \times \! (-i\bn_{j})]_z$
is the operator of the total angular momentum projection.
It can be expressed as  
 $\hat{L}_z  = \sum_j (A^{\dag}_{j}A_{j} -  B^{\dag}_{j}B_{j})$.
Therefore, the basis states (\ref{basis}) have definite total angular 
momentum projections $M_z = n_1 + n_2 - m_1 - m_2$, i.e., are compatible
with the axial symmetry.

The total Hamiltonian $\mathcal{H}$ is characterized by yet another 
geometrical symmetry, namely, magnetic translations (MT).\cite{Zak64}
Indeed, the Hamiltonian commutes with a two-component vector
operator, $[\mathcal{H},\hat{\bf K}]=0$, where 
$\hat{\bf K} =  \sum_j \hat{\bf K}_j$ is the generator of MT
for the whole system.\cite{Sim78,Hir83,Dzy00}
The generators of MT for individual particles are given by
$\hat{\bf K}_j = \hat{\bPi}_j - \frac{q_j}{c} {\bf r}_j \times {\bf B}$.
In the symmetric gauge,
$\hat{\bf K}_j({\bf B}) = \hat{\bPi}_j(-{\bf B})$.
Independent of the gauge, all components of $\hat{\bf K}_j$ and $\hat{\bPi}_j$ commute
with each other: 
$[\hat{K}_{jp},\hat{\Pi}_{jq}]=0$, $p,q=x,y$.
Therefore, the  exact eigenstates of $\mathcal{H}$ can be characterized by
the eigenvalues of the mutually commuting operators $\hat{L}_z$ 
and $\hat{\bf K}^2$. However, the basis (\ref{basis}) is
{\em not\/} compatible with magnetic translations.

%%%%%%%%%%%%%%%%%%%%%%%%%%%%%%%%%
\subsection{Operator of Magnetic Translations (MT)}

To elucidate this, let us indicate the connection between 
the intra-LL raising $B^{\dag}_j$ and lowering  $B_j$
operators with the MT generators $\hat{\bf K}_j$. The components of the
latter are canonically conjugate: 
$[ \hat{K}_{jx} , \hat{K}_{jy} ] = i \tfrac{\hbar B}{c} q_j \neq 0$.
Introducing dimensionless operators 
$\hat{\bf k}_j =   \hat{\bf K}l_j/\hbar$, 
one constructs the raising and lowering intra-LL operators
as $B^{\dag}_j = -i \hat{k}_{j-}$
and $B_j = i \hat{k}_{j+}$. The raising operator
$B^{\dag}({\bf r}_j)$ is given explicitly as a combination of the coordinates 
and derivatives in Eq.~(\ref{lad_em}). We define $\hat{k}_{j\pm}$  
as $\hat{k}_{j\pm} = (\hat{k}_{jx} \pm i \hat{k}_{jy})/\sqrt{2}$.

The components of the MT operator for the whole system, $\hat{\bf K}$, can be expressed as
\begin{eqnarray}
   \label{MTx}
  \hat{K}_x &=&  \hat{K}_{1x} +   \hat{K}_{2x} =
      \frac{i\hbar}{\sqrt{2}}
   \left(    \frac{B_{1}^{\dag} - B_{1}}{l_{1}}
          +  \frac{B_{2}^{\dag} - B_{2}}{l_{2}} \right) \, , \quad\quad \mbox{} \\
   \label{MTy}
  \hat{K}_y &=&  \hat{K}_{1y} +   \hat{K}_{2y} =
      - \frac{\hbar}{\sqrt{2}}
   \left( \frac{B_{1}^{\dag} + B_{1}}{l_{1}}
        +  \frac{B_{2}^{\dag} + B_{2}}{l_{2}}  \right) \, .
\end{eqnarray}
Note now that $\hat{\bf K}^2 = \hat{K}_x^2 + \hat{K}_y^2$ is of the form
\begin{equation}
   \label{MT_K^2}
  \hat{\bf K}^2  =   2\hbar^2 
      \left( \frac{B_{1}^{\dag}B_{1}+\frac{1}{2}}{l_{1}^2} 
           + \frac{B_{2}^{\dag}B_{2}+\frac{1}{2}}{l_{2}^2}
           + \frac{B_{1}^{\dag}B_{2} + B_{2}^{\dag}B_{1}}{l_{1}l_{2}}  \right) \, ,
\end{equation}
i.e., is {\em not\/} diagonal. From here we see that the basis states (\ref{basis}) 
are not eigenstates of  $\hat{\bf K}^2$. 
Therefore, we need to find a transformation that diagonalizes $\hat{\bf K}^2$
while keeping the axial symmetry intact.

%%%%%%%%%%%%%%%%%%%%%%%%%%%%%%%%%%%%%%%%%%%%%%%%%%%%%%%%%%%%%%%%%%%%%%%%%%%%%%%%%
\section{Canonical transformation and new coordinates}
   \label{sec:Can}

To find the transformation that diagonalizes $\hat{\bf K}^2$, 
notice first that the components of $\hat{\bf K}$ commute as
\begin{equation}
        \label{commMT}
  [ \hat{K}_{x} , \hat{K}_{y} ]
               = \frac{i\hbar B}{c} (q_1 + q_2)
               =  \frac{i\hbar B}{c} \, Q   \; ,
\end{equation}
where $Q=q_1 + q_2 >0$ is the negative of the total charge; note the signs.
Let us introduce the dimensionless operator
\begin{equation}
        \label{dimMT}
    \hat{\bf k}
   =   \sqrt{\frac{\hbar c}{QB}} \frac{\hat{\bf K}}{\hbar}
   \equiv  \frac{\hat{\bf K} L_B}{\hbar}
\end{equation}
with the components commuting as canonical variables,
$ [ \hat{k}_x , \hat{k}_y ] = i$; 
$L_B = \sqrt{ \hbar c / QB}$ is the magnetic
length corresponding to the total charge $Q$.
This allows one to introduce the following Bose ladder operators
\begin{eqnarray}
        %\nonumber
          \label{elad1}
 \tilde{B}_{1}^{\dag} &=& -  i \hat{k}_{-}
        = \sqrt{\frac{q_1}{Q}} B^{\dag} 
        + \sqrt{\frac{q_2}{Q}} B^{\dag} 
        \equiv u B_{1}^{\dag}  + v B_{2}^{\dag} \, , \quad\quad \mbox{} \\
%
         %\nonumber
          \label{elad1b}
   \tilde{B}_{1} &=& i \hat{k}_{+}
        = \sqrt{\frac{q_1}{Q}} B 
        + \sqrt{\frac{q_2}{Q}} B   
        \equiv u B_{1}  + v B_{2}   \, ,
\end{eqnarray}
which commute as 
$ [ \tilde{B}_{1} , \tilde{B}_{1}^{\dag} ] = 1$; 
here $\hat{k}_{\pm} = (\hat{k}_x \pm i \hat{k}_y)/\sqrt{2}$ 
(cf.\ with the single-particle operator algebra presented above).
This solves the problem of diagonalizing $\hat{\bf K}^2$. Indeed,  
in the new operators we have
$\hat{k}_{-} = i   \tilde{B}_{1}^{\dag}$
and 
$\hat{k}_{+} = - i  \tilde{B}_{1}$
so that 
$\hat{{\bf k}}^2 =   \hat{k}_{-} \hat{k}_{+} + \hat{k}_{+} \hat{k}_{-} = 
2 \tilde{B}_{1}^{\dag} \tilde{B}_{1}   + 1$.

Note now that we need to consider also the second independent pair of new Bose ladder operators.
These are given by
\begin{eqnarray}
        \label{elad2}
 \tilde{B}_{2}^{\dag} & = & - v B^{\dag}_1 + u B^{\dag}_2  \, , \\
        \label{elad2b}
 \tilde{B}_{2} & = & - v B_1
                            + u B_2  \, .
\end{eqnarray}
The new operators commute as
$[\tilde{B}_{2}, \tilde{B}_{2}^{\dag}]=1$,
$[\tilde{B}_{1}, \tilde{B}_{2}^{\dag}]=0$, and
$[\tilde{B}_{1}, \tilde{B}_{2}]=0$.

The transformation to the new operators is in fact a Bogoliubov canonical 
transformation generated by the unitary operator
(see, e.g., Refs.~\onlinecite{Dzy00})
\begin{eqnarray}
        \label{eBogS}
      S & = &  \exp ( \Phi  \mathcal{M} ) \, , \\
        \label{egenM}
   \mathcal{M} & = &  B^{\dag}({\bf r}_2)  B({\bf r}_1) 
           - B^{\dag}({\bf r}_1) B({\bf r}_2) \, ,
\end{eqnarray}
where $\Phi$ is the transformation parameter and
$u = \cos\Phi=\sqrt{q_1/Q}$,
$v=\sin\Phi =\sqrt{q_2/Q}$.
This transformation can be conveniently presented in the following matrix form:
\begin{eqnarray}
        \label{eB-Brho}
\left( \begin{array}{c}
                   \tilde{B}^{\dag}_1 \\
                   \tilde{B}^{\dag}_2 
                           \end{array} \right) &=&  
   \left( \begin{array}{c}
                   S B^{\dag}({\bf r}_1) S^{\dag} \\
                   S B^{\dag}({\bf r}_2) S^{\dag}
                           \end{array} \right)  
 = \mathsf{G} \left( \begin{array}{c} B^{\dag}({\bf r}_1) \\
                                   B^{\dag}({\bf r}_2)
                 \end{array} \right)  \, , \quad \mbox{}       \\
  \mathsf{G} & \equiv &    \left( \begin{array}{rr}
              u     &   v   \\
            - v     &    u 
                       \end{array} \right)  \,  .
\end{eqnarray}
The matrix $\mathsf{G}$ is an $SU(2)$ matrix, which reflects the fact that the operator 
$\mathcal{M}$ consists of the generators of the $SU(2)$ algebra, 
see Appendices~\ref{Ap:Tr} and \ref{Ap:Dis} and Refs.~\onlinecite{Per86,Gil05}.

%%%%%%%%%%%%%%%%%%%%%%%%%%%%%%%%%%%%%%%%%%%%%%%%%%%%%%%%%%%%%%%%%%%%%%%%%%%%%%%%%
\subsection{New coordinates}

The above canonical transformation introduces, implicitly, 
new effective coordinates
in which one degree of freedom separates. To demonstrate this, let us
write down the coordinate representations of the new operators:
\begin{eqnarray}
        \nonumber 
           %\label{elad_new1}
\mbox{} \hspace{-1pt} \mbox{} \tilde{B}^{\dag}_{1}   &=&  \frac{1}{\sqrt{2}}
  \left[ \frac{1}{2} \left( \frac{u z_1^{\ast}}{l_{1}} 
                          +  \frac{vz_2^{\ast}}{l_{2}}  \right)   \
  \!\! -  2 \left( u l_{1} \frac{\partial \hphantom{,} }{\partial z_1}
                 + v l_{2} \frac{\partial \hphantom{,} }{\partial z_2} \right) \right]  
        % \mbox{} \hspace{6pt}   \mbox{}  
                \; , \\   
       \nonumber    
        %\label{elad_new2}
\mbox{} \hspace{-1pt} \mbox{} \tilde{B}^{\dag}_{2}   &=&  \frac{1}{\sqrt{2}}
  \left[ \frac{1}{2} \left( \frac{u z_2^{\ast}}{l_{2}}  
                          - \frac{v z_1^{\ast}}{l_{1}}  \right)   \
 \!\! -  2 \left( u l_{2} \frac{\partial \hphantom{,} }{\partial z_2}  
                - v l_{1} \frac{\partial \hphantom{,} }{\partial z_1} \right) \right]  \; .  
\end{eqnarray}
This suggests that the complex variables 
$\mathcal{Z}_1^{\ast} = \frac{u}{l_{1}} z_1^{\ast} 
 + \frac{v}{l_{2}} z_2^{\ast}  $
and 
$\mathcal{Z}_2^{\ast} = \frac{u }{l_{2}}^{\vphantom{X}} z_2^{\ast} 
 - \frac{v }{l_{1}}_{\vphantom{x}} z_1^{\ast}$
may be relevant. 
Accordingly, let us introduce the new dimensionless variables using the matrix form:
\begin{equation}
        \label{eF}
             \left( \begin{array}{c} \ro_1   \\
                                     \ro_2   \end{array}  \right)
  =  \mathsf{F} \left( \begin{array}{c} {\bf r}_1  \\
                                    {\bf r}_2  \end{array} \right)
                  \quad , \quad
      \mathsf{F} = \left( \begin{array}{rr}
            \tfrac{u}{l_{1}}  &    \tfrac{v}{l_{2}}  \\
           -\tfrac{v}{l_{1}}^{\vphantom{XX}}   &   \tfrac{u}{l_{2}}^{\vphantom{XX}}
                       \end{array} \right)                      \, .
\end{equation}
The derivatives in the new variables are determined, as usual, by the inverse of 
the transposed matrix, $(\mathsf{F}^{\mathrm{T}})^{-1}$. We obtain
\begin{eqnarray}
        \nonumber
 \left( \begin{array}{c} \tfrac{\partial \hphantom{a} }{\partial\ro_1}   \\
                         \tfrac{\partial \hphantom{a} }{\partial\ro_2}^{\vphantom{XX}}   
          \end{array}  \right)
  &=& \left( \mathsf{F}^{\mathrm{T}} \right)^{-1}
 \left( \begin{array}{c} \tfrac{\partial \hphantom{a} }{\partial{\bf r}_1}  \\
                         \tfrac{\partial \hphantom{a} }{\partial{\bf r}_2}^{\vphantom{XX}} 
        \end{array} \right)  \, , \\
              \label{eFder}   
      \left( \mathsf{F}^{\mathrm{T}} \right)^{-1}
     &=&   \left( \begin{array}{rr}
             u l_{1}   &   v l_{2}   \\
           - v l_{1}   &   u l_{2}
                       \end{array} \right)   \, .
\end{eqnarray}
This shows that indeed we have the derivatives in the new coordinates 
$\mathcal{Z}_1$ and $\mathcal{Z}_2$, so that
$ \tilde{B}^{\dag}_{1}   =   \frac{1}{\sqrt{2}}
      \left( \frac{1}{2} \mathcal{Z}_1^{\ast} - 2 \frac{\partial  \hphantom{Z} }{\partial \mathcal{Z}_1 } 
             \right) $
and 
$ \tilde{B}^{\dag}_{2}   =  \frac{1}{\sqrt{2}}
      \left( \frac{1}{2} \mathcal{Z}_2^{\ast} - 2 \frac{\partial \hphantom{Z} }{\partial \mathcal{Z}_2 } 
              \right) $.

What is the physical meaning of the new coordinates $\ro_1$ and $\ro_2$? 
To answer this question let us note that the matrices 
$\mathsf{F}$, Eq.~(\ref{eF}), and $(\mathsf{F}^{\mathrm{T}})^{-1}$, Eq.~(\ref{eFder}), 
can be presented as
\begin{equation}
        \label{eF_rev}                           
    \mathsf{F} =
        \frac{1}{L_B} \left( \begin{array}{rr}
            u^2  &    v^2  \\
          - uv   &    uv
                       \end{array} \right)   \;\; , \;\;
    \left( \mathsf{F}^{\mathrm{T}} \right)^{-1} =   
        L_B \left( \begin{array}{rr}
                  1     &    1  \\
         - \tfrac{v}{u}   &   \tfrac{u}{v}
                       \end{array} \right)  \,  .
\end{equation}
This allows one to introduce the following coordinates related
to $\ro_1$ and $\ro_2$ by scaling:
\begin{eqnarray}
        \label{eR1}
 {\bf R}_1  &=& L_B \ro_1 = u^2 {\bf r}_1  + v^2 {\bf r}_2 =
   \frac{ q_1 {\bf r}_1 + q_2 {\bf r}_2 }{Q} \; , \\
        \label{eR2}
 {\bf R}_2  &=& L_B \ro_2 = uv ( {\bf r}_2 - {\bf r}_1 ) =
   \frac{\sqrt{q_1 q_2}}{Q} \, ( {\bf r}_2 - {\bf r}_1 )     \; . 
\end{eqnarray}
Using (\ref{eF_rev}) one can show that the transformed 
operators are conveniently expressed in the new variables as
\begin{eqnarray}
        \label{elad_new12}
 \tilde{B}^{\dag}_{1} & =  &  
%   S B^{\dag}({\bf r}_1) S^{\dag} = 
     B^{\dag}({\bf R}_1) 
     =\frac{1}{\sqrt{2}} \left( \frac{Z_1^{\ast} }{2 L_B} -
      2 L_B \frac{\partial \hphantom{Z} }{\partial Z_1} \right)  \; , \\
        \label{elad_new22}
 \tilde{B}^{\dag}_{2} & = &  
%  S B^{\dag}({\bf r}_2) S^{\dag} = 
 B^{\dag}({\bf R}_2) 
     =\frac{1}{\sqrt{2}} \left( \frac{Z_2^{\ast} }{2 L_B} -
      2 L_B \frac{\partial \hphantom{Z} }{\partial Z_2} \right) \, .
\end{eqnarray}
We conclude that the canonical transformation generates the new coordinates, 
which are the ``center-of-charge'' (a weighted dipole) coordinate ${\bf R}_1$ and
a weighted relative coordinate ${\bf R}_2$.

%%%%%%%%%%%%%%%%%%%%%%%%%%%%%%%%%%%%%%%%%%%%%%%%%%%%
\subsection{Do variables separate completely?}

We will see that the above transformation introduces new effective quasiparticles with 
coordinates ${\bf R}_1$ and ${\bf R}_2$. 
Let us first establish the form of the total Hamiltonian in the new variables. 
The interaction Hamiltonian takes the form
\begin{equation}
        \label{Ham_int}
V_{\rm int} = V(|{\bf r}_1 - {\bf r}_2 |) = V\left( \frac{| {\bf R}_2 |}{uv} \right) \; .
\end{equation}
The important feature is that $V_{\rm int}$ {\em does not depend\/} on ${\bf R}_1$
leading to a partial separation of variables.  

At this point it might seem that the coordinate ${\bf R}_1$ can be completely
separated. However, generally, this is not the case. Indeed,
the Hamiltonian of free particles (\ref{Ham_0}) expressed in the new
coordinates ${\bf R}_1$ and ${\bf R}_2$  assumes the form
\begin{eqnarray}
              \nonumber
\mathcal{H}_0 &=& \hbar\tilde{ \omega}_{1} \left( 
            A^{\dag}({\bf R}_1) A({\bf R}_1) + \tfrac{1}{2} \right)  \\
     \nonumber
                                    & & + {\mbox \,\,\,}
            \hbar\tilde{ \omega}_{2} \left( 
            A^{\dag}({\bf R}_2) A({\bf R}_2) + \tfrac{1}{2} \right)   \\
            \label{Ham_0rho}
                   & & +  {\mbox \,\,\,}
                 \hbar\delta\omega  \left( 
            A^{\dag}({\bf R}_1) A({\bf R}_2) + {\rm H.c.} \right)  \; .
\end{eqnarray}
Here the frequencies are given by
\begin{eqnarray} 
         \label{omega_1}
    \tilde{\omega}_{1} &=& \frac{q_1}{Q} \omega_1  + \frac{q_2}{Q} \omega_2  \; ,  \\  
          \label{omega_2}
    \tilde{\omega}_{2} &=& \frac{q_2}{Q} \omega_1  + \frac{q_1}{Q} \omega_2   \; , \\
         \label{omega_d} 
\label{del_omega} 
    \delta\omega & = & \frac{\sqrt{q_1 q_2}}{Q} 
                 \left( \omega_1  - \omega_2 \right)  \; .
\end{eqnarray}
Notice the coupling term proportional to $\delta\omega$ in Eq.~(\ref{Ham_0rho}).
It is this term that precludes one from separating two degrees of freedom corresponding 
to ${\bf R}_1$. The coupling only vanishes in a magnetic field for particles with a constant 
charge-to-mass ratio $q_1/m_1 = q_2/m_2$, so that 
$\delta\omega = \omega_1 - \omega_2 =0$. In this case ${\bf R}_1$ coincides 
with the center-of-mass coordinate and completely separates from the internal motion.\cite{Lau83,Asa98,Kohn61}

%%%%%%%%%%%%%%%%%%%%%%%%%%%%%%%%%
\subsection{A complete basis of states}
   \label{subsec:Compl}

Because of the coupling term (\ref{Ham_0rho}), switching directly to the new variables 
${\bf R}_1$ and ${\bf R}_2$ is not technically advantageous. We will proceed instead by
constructing a complete basis of states which allows one to circumvent that difficulty.
Notice first that the unitary transformation (\ref{eBogS})
does not generate a new vacuum state:
\begin{equation}
        \label{e-vac}
   S | 0 \rangle = | 0 \rangle  \; .
\end{equation}
This is because the generator $\mathcal{M}$ given by Eq.~(\ref{egenM}) is in the normal form.
This should be contrasted with systems of charges of opposite signs, 
where new squeezed vacuum states are generated.\cite{Dzy00,Dzy07}  
The coordinate representations of the normalized vacuum state 
in the two set of coordinates are
\begin{eqnarray}
        \label{e-vac2}
   \langle {\bf r}_1,{\bf r}_2 | 0 \rangle &=&
   \frac{1}{2\pi l_{1} l_{2} }
  \exp\left(
   - \frac{{\bf r}_1^2}{4l_{1}^2}
   - \frac{{\bf r}_2^2}{4l_{2}^2} \right) \; , \\
   \langle {\bf R}_1,{\bf R}_2 | 0 \rangle &=&
   \frac{1}{2\pi L_{B}^2 }
  \exp\left(
   - \frac{{\bf R}_1^2 + {\bf R}_2^2}{4L_{B}^2} \right)  \; .
\end{eqnarray}
We construct a complete orthonormal set of states compatible 
with both axial and magnetic translational symmetries as follows:
\begin{eqnarray}
          \label{basis_new}
   & \mbox{} & |n_1 , n_2, \overline{ m_2, k } \, \rangle  =   
                       \frac{1}{ \left( n_1! \, n_2! \, m_2! \, k! \right)^{1/2} } \times \\
 \nonumber 
      & & \mbox{} \times \left[ A^{\dag}({\bf r}_1) \right]^{n_1} 
                  \left[ A^{\dag}({\bf r}_2) \right]^{n_2} 
                  \left[ B^{\dag}({\bf R}_2) \right]^{m_2} 
                  \left[ B^{\dag}({\bf R}_1) \right]^k | 0 \rangle   \; .
\end{eqnarray}
Notice the mixed character of this representation: the inter-LL operators are in the
old variables ${\bf r}_1$ and ${\bf r}_1$, while the intra-LL operators
are canonically transformed and, as such, are expressed in the new variables ${\bf R}_1$ and ${\bf R}_2$.
The latter is indicated by an overline sign in the notations for the bra and ket states.
Formally, the states (\ref{basis_new}) are the Perelomov coherent states of the $SU(2)$ group;\cite{Per86}
this will be considered in more detail elsewhere.\cite{Dzy_tobe}

Acting by the operators $\hat{L}_z$ and $\hat{\bf K}^2$ on states (\ref{basis_new}) we obtain
\begin{eqnarray}
        \nonumber
   \hat{L}_z |n_1, n_2, \overline{ m_2, k}  \, \rangle  &=& 
  (n_1 + n_2 - m_2 - k) | n_1, n_2, \overline{ m_2, k } \,  \rangle  \; , \\
\label{eigen_LK}
   \hat{\bf K}^2 | n_1, n_2,\overline{ m_2, k } \, \rangle &=& 
                \frac{\hbar^2}{L_B^2} (2k+1)| n_1, n_2, \overline{ m_2, k } \, \rangle   \; .
\end{eqnarray}
We see that the basis states (\ref{basis_new}) are simultaneous eigenstates 
of the generators of rotations $\hat{L}_z$ and 
magnetic translations squared $\hat{\bf K}^2$, i.e.,
are indeed compatible with both the axial 
symmetry and magnetic translations. For identical particles (when, in particular, $q_1 = q_2$) 
the permutational symmetry imposes an additional constraint\cite{Lau83,QHE,QHE1,Dzy00} 
on the parity of $n_2 + m_2$.    

The meaning of the oscillator quantum number $k$ is the 
distance squared to the center of the cyclotron orbit of the complex as a whole.\cite{Sim78,Hir83,Dzy00}
There is the macroscopic Landau degeneracy in $k$ for a charged complex
in ${\bf B}$. 
Because of the degeneracy in $k$, from now on 
we will only consider the $k=0$ states,
$|n_1, n_2, \overline{ m_2, k } \, \rangle \equiv |n_1, n_2, \overline{m_2^{\vphantom{:}}} \, \rangle$. 
Geometrically, these states are at the minimal distance 
from the origin. All other states $k = 1, 2, \ldots$ are displaced
from the origin but, because of the degeneracy, have exactly same energies. 
We consider this in more detail in two sections below.

%%%%%%%%%%%%%%%%%%%%%%%%%%%%%%%%%
\subsection{Guiding center coordinates: Quantum equations of motion}

Let us take a closer look at the quantum dynamics of the degrees of freedom
associated with the introduced transformed operators
$B^{\dag}({\bf R}_1)$ and $B^{\dag}({\bf R}_2)$. 
It is instructive
first to use the connection between the intra-LL operators
of individual particles $B^{\dag}({\bf r}_j)$, $j=1, 2$, and the operators 
$\hat{{\bf r}}_j$ describing positions of their guiding centers in a magnetic field. 
These have the form
$\hat{{\bf r}}_j = - \tfrac{c}{q_j B} \hat{\bf K}_j \times {\bf e}_{z}$,
where ${\bf e}_{z}$ is the unit vector in the magnetic field direction; 
let us remind that we consider negatively charged particles of charges $-q_j < 0$. 
Expressing $\hat{\bf K}_j$ in terms of the intra-LL operators 
gives 
$\hat{x}_j = [B^{\dag}({\bf r}_j) + B({\bf r}_j)]l_j/\sqrt{2}$
and 
$\hat{y}_j = i [B^{\dag}({\bf r}_j) - B({\bf r}_j)]l_j/\sqrt{2}$.
The quantum equations of motion for $\hat{{\bf r}}_j$ have the form
\begin{equation}
        \label{hatr_Ham}
\frac{d \hat{{\bf r}}_j}{d t} = 
   \frac{i}{\hbar} \left[ \mathcal{H}, \hat{{\bf r}}_j \right] = 
        \frac{c}{B q_j} \left( 
            - \frac{\partial V_{\rm int} }{\partial {\bf r}_j} 
                        \right) \quad, \quad j=1, 2 \; .
\end{equation} 
Noticing that $\partial V_{\rm int} / \partial {\bf r}_1 = 
- \partial V_{\rm int} / \partial {\bf r}_2$
and rearranging Eqs.~(\ref{hatr_Ham}) we get 
\begin{eqnarray}
        \label{hatr_Ham2}
     \frac{d \hat{{\bf R}}_1}{d t}   &=&  0  \quad , \quad 
\hat{{\bf R}}_1 = \frac{ q_1 \hat{{\bf r}}_1 + q_2 \hat{{\bf r}}_2 }{Q} \, ,  \\
\frac{d \hat{{\bf R}}_2}{d t}    &=& 
           \frac{c}{B \sqrt{q_1 q_2}}
           \left( - \frac{\partial V_{\rm int} }{\partial {\bf r}_2} \right)
             \times {\bf e}_{z}     \, , \\
                  \label{hatr_Ham3}
   \hat{{\bf R}}_2 &=& \frac{\sqrt{q_1q_2}}{Q}  
            \left( \hat{{\bf r}}_2 - \hat{{\bf r}}_1  \right)  \; , 
\end{eqnarray} 
where $Q=q_1 + q_2$. The operators $ \hat{{\bf R}}_1$ and $ \hat{{\bf R}}_2$ 
are the guiding center operators for the new variables  
${\bf R}_1$ and ${\bf R}_2$ introduced above in Eqs.~(\ref{eR1}) and (\ref{eR2}). 
The former are expressed via the transformed intra-LL operators as 
$\hat{X}_j = [B^{\dag}({\bf R}_j) + B({\bf R}_j)]L_B/\sqrt{2}$
and 
$\hat{Y}_j = i [B^{\dag}({\bf R}_j) - B({\bf R}_j)]L_B/\sqrt{2}$. 
Notice that
the same effective magnetic length $L_B$ is involved in both of them. 

We see that the guiding center operator $ \hat{{\bf R}}_1$ corresponding to the ``center-of-charge'' 
coordinate is a conserved quantity, while the relative coordinate guiding center 
$ \hat{{\bf R}}_2$ experiences a drift motion in the direction perpendicular to the 
gradient of the potential $V_{\rm int}$ and the magnetic field ${\bf B}$.

There is a direct connection between $ \hat{{\bf R}}_1$ and the integral of the 
motion $ \hat{{\bf K}}$:    $ \hat{{\bf R}}_1 = -   L_B^2 \hbar^{-1} \hat{{\bf K}} \times {\bf e}_{z}$.
We see that $ \hat{{\bf R}}_1^2 = [2 B^{\dag}({\bf R}_1) B({\bf R}_1) + 1] L_B^2$. 
Therefore,  quantization of the oscillator quantum number $k=0, 1, \ldots$ 
directly determines the average distances squared of the ``center-of-charge'' from the origin,
$ \langle \hat{{\bf R}}_1^2 \rangle = (2 k + 1) L_B^2$.

%%%%%%%%%%%%%%%%%%%%%%%%%%%%%%%%%
\subsection{Connection with finite Magnetic Translations}

The operator of finite MT for the whole system is\cite{Zak64,Sim78}
\[
\hat{T}({\bf a}) = \exp\left( \frac{i}{\hbar}  {\bf a} \cdot \hat{{\bf K}}   \right) \; .
\]
The group of MTs is non-abelian (non-commutative) for charged systems. We have  
\begin{equation}
           \label{T_ab }
     \hat{T}({\bf b}) \, \hat{T}({\bf a}) = 
     \exp\left(i\frac{ [{\bf a} \times {\bf b}]_z }{2 L_B^2} \right) \hat{T}( {\bf a} + {\bf b} )  \; .
\end{equation}
When expressed in terms of the new coordinates, $\hat{T}({\bf a})$ becomes of the form
\begin{equation}
           \label{T_a }
     \hat{T}({\bf a}) =  \exp\left(i\frac{ [{\bf a} \times {\bf R}_1]_z }{2 L_B^2} \right)
                         \exp\left( {\bf a} \cdot \bn_{{\bf R}_1}  \right)  \; .
\end{equation}
This means that only the ${\bf R}_1$ coordinate is affected by magnetic translations:
\begin{equation}
           \label{T_aPsi } 
            \hat{T}({\bf a}) \Psi({\bf R}_1, {\bf R}_2) = 
                     \exp\left(i\frac{ [{\bf a} \times {\bf R}_1]_z }{2 L_B^2} \right)
                        \Psi({\bf R}_1 + {\bf a}, {\bf R}_2)    \; .
\end{equation}
Essentially, the operator of finite magnetic translations is an oscillator displacement operator\cite{Per86}
\[
\hat{T}({\bf a}) = \exp\left[ \alpha B^{\dag}({\bf R}_1) - \alpha^* B({\bf R}_1) \right]
\]
with $\alpha = -(a_x - i a_y)/\sqrt{2}L_B$.

\begin{widetext}
%%%%%%%%%%%%%%%%%%%%%%%%%%%%%%%%%
\section{New quasiparticles and the unitary transformation of the Hamiltonian}

The interaction Hamiltonian is block-diagonal in the constructed basis:
\begin{equation}
      \label{Hint_mat}  
       \langle n_1', n_2', \overline{ m_2', k' } \, |
                 V_{\rm int} |n_1, n_2, \overline{ m_2^{\vphantom{a}} , k } \, \rangle    
    =  \delta_{k,k'} \, \delta_{n_1 + n_2 - m_2 - k,n_1' + n_2' - m_2' - k'} \,
   \langle n_1', n_2', \overline{m_2'} \, | 
                   V_{\rm int} |n_1, n_2, \overline{m_2^{\vphantom{a}}} \, \rangle   \; .
\end{equation}
\end{widetext}
Note that the matrix elements (\ref{Hint_mat}) do not depend on the oscillator 
quantum number $k$ and can be calculated using the $k=0$ states defined 
above in Sec.~\ref{subsec:Compl} as $|n_1, n_2, \overline{m_2^{\vphantom{,}}} \, \rangle $. 

It is now convenient to perform the transformation of the inter-LL operators
to the variables ${\bf R}_1$ and ${\bf R}_2$. This can be done as follows:
\begin{eqnarray}
        \label{eA-Brho}
    &\mbox{}&               \left( \begin{array}{c} A^{\dag}({\bf r}_1) \\
                                   A^{\dag}({\bf r}_2)
                 \end{array} \right) = 
    \left( \begin{array}{c}
  U A^{\dag}({\bf R}_1) U^{\dag} \\
                    U A^{\dag}({\bf R}_2) U^{\dag} 
                           \end{array} \right)  = \\
 & & = \mathsf{G}^{-1} \left( \begin{array}{c} A^{\dag}({\bf R}_1) \\
                                   A^{\dag}({\bf R}_2)
                 \end{array} \right)      \quad , \quad                            
 \mathsf{G}^{-1} =     \left( \begin{array}{rr}
             u     &   - v   \\
             v     &    u 
                       \end{array} \right)  \;  .
\end{eqnarray} 
This is a canonical transformation similar to the inverse of 
the transformation of the intra-LL operators (\ref{eB-Brho}). 
Accordingly, the unitary operator involved is 
\begin{eqnarray}
        \label{BogU}
      U & = &  \exp ( - \Phi \mathcal{L} ) \; , \\
        \label{egenL}
  \mathcal{L} & = &  A^{\dag}({\bf R}_2) A({\bf R}_1) 
           - A^{\dag}({\bf R}_1) A({\bf R}_2) \; ,
\end{eqnarray}
cf.\ Eq.~(\ref{eBogS}).
Notice that $[ U, \mathcal{L}] =0$ so that the antihermitian
generator $\mathcal{L}$  has the same form in both sets of coordinates:
$\mathcal{L} = U  \mathcal{L} U^{\dag} = 
A^{\dag}({\bf r}_2) A({\bf r}_1) - A^{\dag}({\bf r}_1) A({\bf r}_2)$.

Performing the transformation we obtain matrix elements (\ref{Hint_mat}) as
\begin{eqnarray}
   \label{Hint_mat2} 
    \langle n_1', n_2', \overline{m_2'} \, | 
                \, V_{\rm int} \, 
                          |\, n_1, n_2, \overline{m_2^{\vphantom{a}}} \, \rangle =   
                \mbox{}\hspace{10pt} \mbox{} \\
           \label{Hint_matU}
  =  \langle \, \overline{ n_1', n_2', m_2'} \, | 
                \,  U^{\dag} V_{\rm int} U \, 
                      | \, \overline{n_1, n_2,  m_2^{\vphantom{a}}}  \, \rangle    \;  .
\end{eqnarray}
Here we introduced the states that are expressed solely in terms of the new variables: 
\begin{eqnarray}
        %\nonumber 
      \label{basis_neww}     
  | \, \overline{n_1, n_2,  m_2^{\vphantom{a}}} \, \rangle & = &  
U^{\dag} \, | n_1, n_2, \overline{m_2^{\vphantom{a}}} \, \rangle =  
               \mbox{} \quad \mbox{} \\
   \nonumber     
                 \frac{1}{ \left( n_1! \, n_2! \, m_2! \right)^{1/2} } 
                                     &\mbox{}& \mbox{}\hspace{-20pt} \mbox{}
                  \left[ A^{\dag}({\bf R}_1) \right]^{n_1} 
                  \left[ A^{\dag}({\bf R}_2) \right]^{n_2} 
                  \left[ B^{\dag}({\bf R}_2) \right]^{m_2} 
                   | 0 \rangle   \; .
\end{eqnarray}
We used the fact that $U^{\dag} |0 \rangle = |0 \rangle $, which is similar to Eq.~(\ref{e-vac}).
We arrive therefore to the desired representation in the new variables ${\bf R}_1$, 
${\bf R}_2$. 

In this representation, the Hamiltonian undergoes the unitary transformation 
\begin{equation}
  \label{can_H}
    \mathcal{H}  \rightarrow \overline{\mathcal{H}} = U^{\dag} \mathcal{H} U \; .
\end{equation}
For the interaction Hamiltonian, the change   
$V_{\rm int} \rightarrow \overline{V}_{\rm int} = U^{\dag} V_{\rm int} U$
is evident from Eq.~(\ref{Hint_matU}).
Transformation of the free Hamiltonian $\mathcal{H}_{0} \rightarrow \overline{\mathcal{H}}_{0}=
U^{\dag} \mathcal{H}_{0}  U $
yields 
\begin{eqnarray}
         \nonumber
\overline{\mathcal{H}}_{0} & = & 
     \hbar\omega_{1} \left( 
            A^{\dag}({\bf R}_1) A({\bf R}_1) + \tfrac{1}{2} \right)  \\
   \label{Ham_0U} 
  & &  \mbox{} + 
            \hbar\omega_{2} \left( 
            A^{\dag}({\bf R}_2) A({\bf R}_2) + \tfrac{1}{2} \right)  \; .
\end{eqnarray}
The change of coordinates follows from (\ref{eA-Brho}).
Now we see that indeed new quasiparticles with coordinates ${\bf R}_1$ and ${\bf R}_2$,
the {\em same\/} cyclotron frequencies $\omega_{1}$ and $\omega_{2}$ [cf.\ Eq.~(\ref{Ham_0rho})],
and a {\em modified\/} interaction 
$\overline{V}_{\rm int} = U^{\dag} V_{\rm int} U$ 
emerge in the developed formalism.

Note that the states $| \, \overline{n_1, n_2,  m_2^{\vphantom{a}}} \, \rangle $ 
depend only on three quantum numbers
corresponding to the three remaining degrees of freedom. Note also that while 
$V_{\rm int}$ depends only on ${\bf R}_2$, the transformed interaction Hamiltonian 
$\overline{V}_{\rm int} = U^{\dag} V_{\rm int} U$ 
depends on both ${\bf R}_1$ and 
${\bf R}_2$. This reflects the fact that, generally, no complete separation of variables 
is possible in a magnetic field. 

The unitary transformation (\ref{can_H}) together with simultaneous 
diagonalization of the integrals of the motion $\hat{{\bf K}}^2$ and $\hat{L}_z$ 
(Sec.~\ref{sec:Can}) are the main formal results of this paper.    
Two different techniques for calculating the interaction 
matrix elements in higher Landau levels are presented
in Appendices~\ref{Ap:Tr} and \ref{Ap:Untr}. 
These are connected with two different representations of the $SU(2)$ group.\cite{Gil05}
We discuss the relevant $SU(2)$ algebras in Appendix~\ref{Ap:Dis}.
Below we concentrate on eigenstates and eigenenergies in lowest Landau levels.

%%%%%%%%%%%%%%%%%%%%%%%%%%%%%%%%%
\subsection{Eigenfunctions and eigenvalues in zero Landau level}

Let us denote the states 
$| \, \overline{n_1, n_2,  m^{\vphantom{a}}} \, \rangle$ in zero LLs $n_1=0$, $n_2=0$  
and with oscillator quantum number $k=0$ as  
\begin{eqnarray}
        \label{e-basis}
    | \, \overline{m^{\vphantom{:}}} \, \rangle & = &  \frac{1}{\sqrt{m!}}
         \left[ B^{\dag}({\bf R}_2) \right]^m | 0 \rangle  \; .
\end{eqnarray}
These states form a complete orthonormal basis of states compatible 
with both  axial and translational symmetries in the lowest LL.
The coordinate representation is given by 
\begin{equation}
        \label{m-states}
   \langle {\bf R}_1,{\bf R}_2 | \, \overline{m^{\vphantom{:}}} \, \rangle =
   \frac{1}{2\pi L_{B}^2 \sqrt{m!}}  \left( \frac{Z_2^*}{\sqrt{2} \, L_B} \right)^m
   \mathrm{e}^{- \tfrac{{\bf R}_1^2 + {\bf R}_2^2}{4 L_{B}^2} } \; , 
\end{equation}
i.e., has a standard form of the factored electron wavefunctions in zero LL.\cite{QHE,QHE1,Dzy00}

Since $U$ has a normal form, the states in zero LLs do not change 
under the action of this operator: 
$U | \, \overline{m^{\vphantom{a}}}  \, \rangle = | \, \overline{m^{\vphantom{a}}} \, \rangle$.
Therefore, 
\begin{equation}
        \label{Vm_zero}
  \langle \, \overline{m'}  \, | \overline{V}_{\rm int} | \, \overline{m^{\vphantom{a}}} \, \rangle =
       \langle \,\overline{m'} \,| V_{\rm int} | \,\overline{m^{\vphantom{a}}}\, \rangle = 
\delta_{m',m} \langle \,\overline{m^{\vphantom{a}}} \,| V_{\rm int} 
| \,\overline{m^{\vphantom{a}}} \,\rangle \, .
\end{equation}
We used the fact that the interaction Hamiltonian   
$V_{\rm int} = V\left( \frac{| {\bf R}_2 |}{uv} \right)$, Eq.~(\ref{Ham_int}), 
possesses the axial symmetry and, therefore, does not have off-diagonal 
matrix elements in this basis. 
This allows one to calculate the energy eigenvalues in a given LL as the expectation 
values of the interaction Hamiltonian. This is similar to the case of the electron system
in a strong magnetic field and determines the Haldane pseudopotentials.\cite{Lau83,QHE1,QHE,Hal87}

A relatively wide class of pairwise interaction 
potentials with combined power-law and Gaussian distance dependencies  
can be treated analytically:
\begin{equation}
        \label{eUint_G}
 V(|{\bf r}_1 - {\bf r}_2|) = \frac{W_0} { |{\bf r}_1 - {\bf r}_2|^{\epsilon} } \,
               \exp\left( - \frac{|{\bf r}_1 - {\bf r}_2|^2}{2L^2} \right)  \; .
\end{equation}
Here $W_0$ is the potential  amplitude, 
$\epsilon$ determines the power-law dependence, and $L$ is the characteristic 
length. Such effective potentials may be relevant for a description
of interactions between quasiparticles in the fractional quantum Hall effect regime.\cite{Jain05}

Switching from the relative coordinate $|{\bf r}_1 - {\bf r}_2|$ 
to the variable $R_2 = |{\bf R}_2|$ from Eq.~(\ref{eR2}) and calculating the diagonal 
matrix elements using the coordinate representation (\ref{m-states})
gives the eigenenergies of the problem (Haldane pseudopotentials).
These have the following form: 
\begin{eqnarray}
        \nonumber
   V_{0,m} &=& \langle \,\overline{m^{\vphantom{a}}} \,| V_{\rm int} | \,\overline{m^{\vphantom{a}}} \, \rangle  \\
        \nonumber      
   V_{0,m} &=& \frac{W_0}{L_B^{\epsilon} }  \left( \frac{q_1q_2}{Q^2} \right)^{\tfrac{\epsilon}{2}} 
               \left[ 1 + \frac{Q^2}{q_1 q_2 } \left(\frac{L_B}{L} \right)^2    
                     \right]^{ \tfrac{\epsilon}{2} - m - 1 }  \times                      \\
                    \label{eUm_G}  
	&&  \mbox{} \hspace{20pt} \mbox{} \times 
     \frac{\Gamma(m - \frac{\epsilon}{2} + 1)}{2^{\epsilon/2} m!}  \; ,
\end{eqnarray}
where $\Gamma (x)$ is the gamma-function. 
The eigenenergies $V_{0,m}$ for several sets of the parameters are shown in Fig.~1. 
%%%%%%%%%%%%%%%%%%%%%%%%%%%%%%%%%%%%%%%%%%%%%%%%%%%%%%%%%%%%%%%%%%%%%%%%%%%%%%%%%%%%%%%%
\begin{figure}[t]
\includegraphics[scale=0.95]{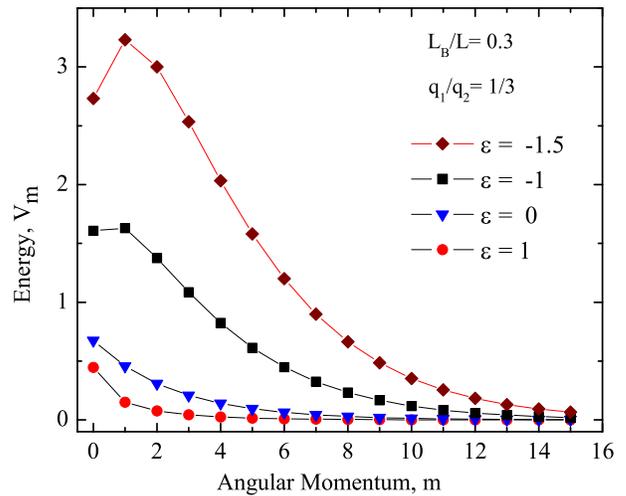}
\caption{\label{fig:VM}
(Color online) 
Eigenenergies (\protect\ref{eUm_G}) for several different power-law dependencies $\epsilon$.
The case $\epsilon=0$ corresponds to a Gaussian interaction potential, 
while  $\epsilon=1$ corresponds to the modified Coulomb 
potential $V_{\rm int} \sim \tfrac{1}{r}\exp(-r^2/L^2)$, see Eq.~(\protect\ref{eUint_G}).
All eigenenergies are given in units of $W_0/L_B^{\epsilon}$ 
for the same value of $L_B/L = 0.3$. The ratio of the charges $q_1/q_2= \frac{1}{3}$.    
 }
\end{figure}
%%%%%%%%%%%%%%%%%%%%%%%%%%%%%%%%%%%%%%%%%%%%%%%%%%%%%%%%%%%%%%%%%%%%%%%%%%%%%%%%%%%%%%%%
The asymptotic behavior $m \gg 1$  of the eigenenergies (\ref{eUm_G})  
is exponential for finite $L$: $V_m \sim x^{-m}m^{-\epsilon/2}$, 
where $x=1 + Q^2L_B^2/q_1 q_2L^2$.
For pure power-law potentials $V_{\rm int} \sim r^{-\epsilon}$  (corresponding 
to $L \rightarrow \infty$), the asymptotics is a power-law $V_m \sim m^{-\epsilon/2}$. 
Note that even in case of a repulsive interaction between the particles,
all states are bound and the eigenspectra are completely discrete.

For $\epsilon = 1$, $L \rightarrow \infty$ and amplitude $W_0 = q_1 q_2$, 
Eq.~(\ref{eUm_G}) gives results for the Coulomb interaction potential 
\begin{equation}
        \label{eUintC}
 V_{\rm int}^{C} 
             =  \frac{q_1q_2}{|{\bf r}_1 - {\bf r}_2|} 
             =  \frac{\sqrt{q_1q_2}}{Q} \, \frac{q_1q_2}{R_2} \; .              
\end{equation}
The eigeneregies in zero LL in this case are given by 
\begin{equation}
        \label{eUm_newC}
   V_{0,m}^{C} = \langle \,\overline{m^{\vphantom{a}} } \,| V_{\rm int}^{C} 
                             | \,\overline{m^{\vphantom{a}} } \, \rangle =
	 \frac{q_1 q_2}{ L_B} \, \frac{ \sqrt{q_1 q_2} }{Q} \, 
	       \frac{\Gamma(m+\frac{1}{2})}{\sqrt{2} \, m!}  \; .
\end{equation}
For $q_1 = q_2$ these reproduce to the Haldane pseudopotentials 
in zero LL.\cite{QHE,Hal83,Hal87,Jain05}

%%%%%%%%%%%%%%%%%%%%%%%%%%%%%%%%%
\subsection{Eigenfunctions and eigenvalues in first Landau levels}
    \label{subsec:First}

We will illustrate how one can obtain the eigenenergies in the few first LLs
on the example of the Coulomb potential. 
To calculate the interaction matrix elements we will use representation (\ref{Hint_mat2}). 
Consider, e.g., the state 
$|\, 1, 0, \overline{m^{\vphantom{a}}} \, \rangle = A^{\dag}({\bf r}_1)|\,0 , 0, \overline{m^{\vphantom{a}}} \, \rangle$.   
Using transformation (\ref{eA-Brho}), 
$A^{\dag}({\bf r}_1) = u A^{\dag}({\bf R}_1) -v A^{\dag}({\bf R}_2)$,
one obtains
\begin{equation}
|\, 1, 0, \overline{m^{\vphantom{a}}} \, \rangle = u |\, \overline{1, 0, m^{\vphantom{a}}} \, \rangle 
- v |\, \overline{0, 1, m^{\vphantom{a}}} \, \rangle \; . 
\end{equation}
The total energy of this state
is given as an expectation value of the total Hamiltonian and is equal to
\begin{equation}
        \label{eUm_new1}
    \langle \, 1, 0, \overline{m^{\vphantom{a}}} \,| \mathcal{H}_0 + V_{\rm int}^{C} 
                       | \,1, 0, \overline{m^{\vphantom{a}}} \, \rangle =
	  \tfrac{3}{2}\hbar\omega_1 + u^2 V_{0,m}^{C} +v^2 V_{1,m}^{C} \; .
\end{equation}
Here $V_{0,m}^{C}$ is given by Eq.~(\ref{eUm_newC}) and 
$V_{1,m}^{C} = \tfrac{4m-1}{4m-2} V_{0,m}^{C}$ are the Coulomb pseudopotentials in the first LL.
 
Analogously, for a state 
$|\, 0, 1, \overline{m^{\vphantom{a}}} \, \rangle = 
A^{\dag}({\bf r}_2)|\,0 , 0, \overline{m^{\vphantom{a}}} \, \rangle$
using 
$A^{\dag}({\bf r}_2) = v A^{\dag}({\bf R}_1) + u A^{\dag}({\bf R}_2)$ from (\ref{eA-Brho})
one gets 
\begin{equation}
|\, 0, 1, \overline{m^{\vphantom{a}}} \, \rangle = v |\, \overline{1, 0, m^{\vphantom{a}}} \, \rangle 
+ u |\, \overline{0, 1, m^{\vphantom{a}}} \, \rangle \; . 
\end{equation}
Therefore, 
\begin{equation}
        \label{eUm_new2}
    \langle \, 0, 1, \overline{m^{\vphantom{a}}} \,| \mathcal{H}_0 + V_{\rm int}^{C} 
     | \, 0 , 1, \overline{m^{\vphantom{a}}} \, \rangle =
	  \tfrac{3}{2}\hbar\omega_2 + v^2 V_{0,m}^{C} + u^2 V_{1,m}^{C} \; .
\end{equation}
Notice that both zero- and first-LL pseudopotentials are involved in Eqs.~(\ref{eUm_new1}) and (\ref{eUm_new2}).

In conclusion, we introduced the operator formalism for partial separation of degrees 
of freedom for complexes consisting of two like but unequal charges in a magnetic field
in the planar geometry. The exact geometrical symmetry, magnetic translations, 
played the central role in our approach. 
We established the connection of the developed formalism with the $SU(2)$ algebras. 
We found analytically Haldane pseudopotentials in the first few Landau levels 
for a class of interaction potentials with combined power-law and gaussian dependencies. 
Our results may be useful for considering Coulomb correlations in Landau levels.

%\mbox{}
%\vspace{-25pt}
%\mbox{}

\section*{Acknowledgments}

ABD is grateful to the Kavli Institute for Theoretical Physics, UC Santa Barbara, 
and to the Max Planck Institute for the Physics of Complex Systems, Dresden,  
for hospitality. This work was supported in part by NSF grants 
DMR-0203560 and DMR-0224225
and by a College Science Award of Cottrell Research Corporation.

\appendix

%%%%%%%%%%%%%%%%%%%%%%%%%%%%%%%%%%%%%%%%%%%%%%%%%%%%%%%%%%%%%%
\section{Matrix elements in higher Landau levels:
          Transformed basis}
                     \label{Ap:Tr}

We present here a derivation of the interaction matrix elements in arbitrary LLs
using the canonically transformed representation (\ref{Hint_matU}).
We first disentangle the operators in the exponent of $U$, Eq.~(\ref{BogU}),
to obtain the normal-ordered form
\begin{eqnarray}
      \label{Sexp}
\mbox{} \hspace{-10pt} \mbox{} & U &  = 
   \exp \left( - \tan \! \Phi \, A^{\dag}_2  A _1 \right)  \times  \\
      \nonumber
  & \times  & \exp \left[  \ln ( \cos\Phi ) 
                  \left( A^{\dag}_1  A_1 
                       - A^{\dag}_2  A_2 \right) \right] \exp\left( \tan \! \Phi \, A^{\dag}_1  A_2 \right) \; ,
\end{eqnarray}
where $\cos\Phi=u=\sqrt{q_1/Q} > 0$.
Expression (\ref{Sexp}) is an analog of the Baker-Campbell-Hausdorff formulas
for the $SU(2)$ group\cite{Per86,Gil05} and is discussed in Appendix~\ref{Ap:Dis} below.
Acting by $U $ on one of a basis states (\ref{basis_neww}) we get 
\begin{widetext}
\begin{eqnarray}
               \label{expandU}
    U |\, \overline{n_1, n_2, m_2^{\vphantom{a}}} \, \rangle  &=& 
    \sum_{k=0}^{n_2} \, \sum_{l=0}^{n_1 + k} \, H_{n_1  n_2}(k,l)
    |\, \overline{n_1+ k - l, n_2 - k + l, m_2^{\vphantom{a}}} \, \rangle \; ,  \\
          \nonumber     
          %\label{Hnn}
  H_{n_1  n_2}(k,l) & = &  u^{n_1-n_2} (-1)^{l} u^{k-l} v^{k+l}
                     \left[  \binom{l}{n_2-k+l} \binom{l}{n_1+k} \right]^{1/2} \,    
                     \left[  \binom{k}{n_1+k}     \binom{k}{n_2} \right]^{1/2} \; .
\end{eqnarray}
%\end{widetext}
Using the fact that $V_{\rm int}=V(R_2/uv)$ does not depend on ${\bf R}_1$, we obtain 
%\begin{widetext}
\begin{eqnarray}
                  \nonumber
&\mbox{}&  \langle \, \overline{n_1', n_2', m_2'} \, | 
                \, U^{\dag} V_{\rm int} U \, 
                          |\, \overline{n_1, n_2, m_2^{\vphantom{a}}} \, \rangle =  
                         \delta_{n_1+n_2 - m_2,n_1'+n_2' - m_2'} \times \\ 
                  \label{matApT}
  && \mbox{} \times \sum_{k'=0}^{n_2'} \, \sum_{l'=0}^{n_1'+k'}
   \sum_{k=0}^{n_2}   \, \sum_{l=0}^{n_1+k} \, \delta_{n_1+k-l,n_1'+k'-l'}
                    H_{n_1'  n_2'}(k',l') H_{n_1  n_2}(k,l) 
                    V_{n_2 - k + l \:\: m_2}^{n_2' - k'+ l' \:\: m_2'} \; .
\end{eqnarray}
\end{widetext}
Here $V_{n \, m}^{n'\, m'} = \delta_{n+m,n'+m'} F_{n \, m}^{n'\, m'}$
is a single-particle matrix element defined, e.g., in 
Eqs.~(43) and (44) of Ref.~\onlinecite{Dzy00}.

%%%%%%%%%%%%%%%%%%%%%%%%%%%%%%%%%%%%%%%%%%%%%%%%%%%%%%%%%%%%%%
\section{Disentangling the operators: the SU(2) algebra}
                     \label{Ap:Dis}

Equation~(\ref{Sexp}) is the analog of the 
Baker-Campbell-Hausdorff formulas for the $SU(2)$ group.\cite{Per86,Gil05}  
It can be obtained as follows. Notice first that operators 
$A^{\dag}_2  A_1 \equiv \mathcal{K}_{-}$, 
$A^{\dag}_1  A_2 \equiv \mathcal{K}_{+}$,
and 
$\frac{1}{2}(A^{\dag}_1  A_1 - A^{\dag}_2  A_2)  \equiv \mathcal{K}_{0}$ 
realize a two-mode representation of the $SU(2)$ algebra 
and satisfy the commutation relations
\begin{equation}
                    \label{SU2}
           [  \mathcal{K}_{0} , \mathcal{K}_{\pm} ]  =  \pm \mathcal{K}_{\pm}  
                                               \quad , \quad 
              [ \mathcal{K}_{-}  , \mathcal{K}_{+}  ]    =   - 2 \mathcal{K}_{0} \; .
\end{equation}

We will note here in passing that the quadratic Casimir operator, i.e., 
the operator commuting with all $SU(2)$ generators 
$\mathcal{K}_{\pm}$ and $\mathcal{K}_{0}$, is of the form
\[
C_2 = \mathcal{K}_{0}^2 +\frac{1}{2}\left(\mathcal{K}_{-}\mathcal{K}_{+} + \mathcal{K}_{+}\mathcal{K}_{-} \right)
= \hat{N}(\hat{N} + 1) \; ,
\]
where, in our case, 
$\hat{N}=\frac{1}{2}( \hat{N}_1 + \hat{N}_2) = \frac{1}{2}( A^{\dag}_1  A_1 + A^{\dag}_2  A_2)$. 
We notice that all states generated by transformation (\ref{expandU}) belong to the same
representation of the $SU(2)$ group and are characterized by the 
same Bargmann index $N= (n_1 + n_2)/2$. Here $n_1$ and $n_2$ are the Landau level numbers. 
The Casimir operator $C_2$ for the $SU(2)$ group is the analog of the total angular momentum operator
${\bf J}^2$ so that $N$ is the analog of the total angular momentum~$J$. 

Notice now that the unitary transformation operator
$U =\exp \left[ 
           \Phi \, \left( \mathcal{K}_{+} - \mathcal{K}_{-}         
                    \right) \right]$,
Eq.~(\ref{BogU}), 
involves two generators of the $SU(2)$ group in the exponent. 
Although various useful representations for such operators can be found in the literature,\cite{Per86,Gil05}
we will present here a derivation of the form (\ref{Sexp}) directly suitable for our purposes.  
In order to disentangle the operators, we will closely follow the method of Ref.~\onlinecite{Kir67}.
Let us first present $U$ in the form
\begin{equation}
   \label{factor1}
    U =   \exp \left( \alpha  \mathcal{K}_{-} \right) 
                       \exp \left( \beta  \mathcal{K}_{0} \right)
                       \exp \left( \gamma  \mathcal{K}_{+} \right)    \; ,
\end{equation}
where $\alpha$, $\beta$, and $\gamma$ are unknown functions of $\Phi$. 
Let us establish the differential equations that these functions obey.
To this end we differentiate Eq.~(\ref{factor1}) with respect
to $\Phi$ and then multiply both parts from right by 
$\exp \left( - \gamma  \mathcal{K}_{+} \right)$, 
$\exp \left( - \beta  \mathcal{K}_{0} \right)$, and 
$\exp \left( - \alpha  \mathcal{K}_{-} \right)$ in the indicated order.
This gives 
\begin{eqnarray}
       \nonumber
       \mathcal{K}_{+}  -  \mathcal{K}_{-}  
     & = &  \alpha' \, \mathcal{K}_{-}  +
     \beta' \, e^{ \alpha \mathcal{K}_{-}} \mathcal{K}_{0}  e^{- \alpha \mathcal{K}_{-}}  \\
      \label{factor2}
       & &  \mbox{} + \gamma' \, e^{ \alpha \mathcal{K}_{-}} e^{ \beta \mathcal{K}_{0}} 
            \mathcal{K}_{0} e^{ - \beta \mathcal{K}_{0}} e^{- \alpha \mathcal{K}_{-}}  \; ,
\end{eqnarray}
where $\alpha'$, $\beta'$, and $\gamma'$ are the derivatives with respect to $\Phi$.
We identify in Eq.~(\ref{factor2}) the similarity transformations of the $SU(2)$ group. 
These can be found by differentiating with respect to the transformation
parameter. We have
\begin{eqnarray}
       \label{Ga}
     G(\alpha) & \equiv & e^{ \alpha \mathcal{K}_{-}} \mathcal{K}_{0}  e^{- \alpha \mathcal{K}_{-} } \; , \\
\nonumber
  \frac{d G}{d \alpha}  & = & e^{ \alpha \mathcal{K}_{-}} [ \mathcal{K}_{-}, \mathcal{K}_{0} ]  
                        e^{- \alpha \mathcal{K}_{-} } = \mathcal{K}_{-} \; , \\
\nonumber
        G(0) &=&  \mathcal{K}_{0} \; \\
 \label{factor3}
     G(\alpha) & = & \alpha \mathcal{K}_{-} + \mathcal{K}_{0}  \; .
\end{eqnarray}
Analogously,
\begin{eqnarray}
       \label{Fb}
     F(\beta) &\equiv&  e^{ \beta \mathcal{K}_{0}}  \mathcal{K}_{+}    e^{ - \beta \mathcal{K}_{0}} \; ,  \\
\nonumber
  \frac{d F}{d \beta} &=& e^{   \beta \mathcal{K}_{0}} 
                                       [ \mathcal{K}_{0}, \mathcal{K}_{+} ]  
                          e^{ - \beta \mathcal{K}_{0}} = F  \; ,  \\
\nonumber
        F(0) &=&  \mathcal{K}_{+}  \; , \\
  \label{Fbf}
    F(\beta) &=& e^{\beta} \mathcal{K}_{+}  \; .
\end{eqnarray}
Finally,
\begin{eqnarray}
      \label{Ja}
     J(\alpha) &\equiv&   e^{ \alpha \mathcal{K}_{-}} \mathcal{K}_{+}  e^{- \alpha \mathcal{K}_{-} } \; ,\\ 
\nonumber
  \frac{d J}{d \alpha} &=& -2 e^{ \alpha \mathcal{K}_{-}} 
                                                      \mathcal{K}_{0} 
                            e^{- \alpha \mathcal{K}_{-} }   \quad , \quad 
  \frac{d^2 J}{d \alpha^2} = -2 \mathcal{K}_{-}   \; , \\
\nonumber
   J(0) &=&  \mathcal{K}_{+}   \quad , \quad  \left. \frac{d J}{d \alpha} \right|_{\alpha=0} = - 2 \mathcal{K}_{0}  \; , \\
     \label{factor6}
       J(\alpha) &=&  - \alpha^2 \mathcal{K}_{-} - 2\alpha \mathcal{K}_{0} + \mathcal{K}_{+}\, .
\end{eqnarray}
We can now identify the linearly independent terms in Eq.~(\ref{factor2}) 
and establish a system of coupled differential equations
\begin{eqnarray}
\nonumber
   && \gamma' e^{\beta} = 1     \; , \\
\nonumber
    && \alpha' + \beta' \, \alpha -\alpha^2 = -1 \; , \\
\nonumber
     &&  \beta' - 2 \alpha = 0    \; .  
\end{eqnarray}
Solutions satisfying the initial conditions
$\alpha(0)=\beta(0)=\gamma(0)=0$ are given by
$ \alpha = - \gamma = \tan \Phi$ and $\beta = 2 \ln( | \cos \Phi |)$.
This gives 
\begin{eqnarray}
   \nonumber
   & \mbox{}& \exp\left[ \Phi \left( \mathcal{K}_{+} - \mathcal{K}_{-}\right)  \right] 
                 =  \exp \left( - \tan\Phi \,\, \mathcal{K}_{-} \right) \times  \mbox{} \quad \quad \quad \mbox{} \\
    \label{K_dis}  
            && \mbox{}\quad \mbox{} \times  \exp \left( 2 \ln(| \cos\Phi |) \,\, \mathcal{K}_{0} \right) 
               \exp \left(  \tan\Phi \,\, \mathcal{K}_{+} \right)   \; ,   
\end{eqnarray}
and Eq.~(\ref{Sexp}) follows.

%%%%%%%%%%%%%%%%%%%%%%%%%%%%%%%%%%%%%%%%%%%%%%%%%%%%%%%%%%%%%%
\section{Matrix elements in higher Landau levels: 
              Initial basis}
                     \label{Ap:Untr}

We present here a derivation of the interaction matrix elements in arbitrary LLs
based on the untransformed representation (\ref{Hint_mat2}). This is similar 
to our treatment in Sec.~\ref{subsec:First} above.     
Using transformation (\ref{eA-Brho}) to express $A^{\dag}({\bf r}_1)$ and  
$A^{\dag}({\bf r}_2)$ in terms of $A^{\dag}({\bf R}_1)$ and $A^{\dag}({\bf R}_2)$
and applying the binomial expansion, 
we obtain the ket states as
\begin{widetext}
\begin{eqnarray}
                \label{expand}
|\, n_1, n_2, \overline{m_2^{\vphantom{,}}} \, \rangle &=& 
\sum_{k_1=0}^{n_1} \, \sum_{k_2=0}^{n_2} \, G_{n_1  n_2}(k_1,k_2)  
              |\, \overline{k_1 + k_2, n_1+n_2-k_1-k_2, m_2^{\vphantom{a}}} \, \rangle \; , \\ 
   \nonumber           
   % \label{F}
   G_{n_1  n_2}(k_1,k_2) &=& (-1)^{n_1 - k_1} \,  u^{n_2-k_2+k_1} v^{n_1-k_1+k_2} \, 
           \binom{k_1}{n_1} \binom{k_2}{n_2} \, 
           \left[ \frac{(k_1+k_2)! (n_1+n_2-k_1-k_2)!}{n_1!n_2!} \right]^{1/2} \; .
\end{eqnarray}
Using the fact that $V_{\rm int}=V(R_2/uv)$ does not depend on ${\bf R}_1$, we obtain
\begin{eqnarray}
                  \nonumber
&\mbox{}&  \langle n_1', n_2', \overline{m_2'} \, | 
                \, V_{\rm int} \, 
                          |\, n_1, n_2, \overline{m_2^{\vphantom{a}}} \, \rangle = 
  \delta_{n_1+n_2 - m_2,n_1'+n_2' - m_2'}  \times \\
        \label{matAp1}\
& & \mbox{} \times \sum_{k_2'=0}^{n_2'}
 \sum_{k_1'=0}^{n_1'} \,  
 \sum_{k_1=0}^{n_1}   \, 
 \sum_{k_2=0}^{n_2}   \, \delta_{k_1+k_2,k_1'+k_2'}
   G_{n_1'  n_2'}(k_1',k_2') G_{n_1  n_2}(k_1,k_2) 
                    V_{n_1+n_2-k_1-k_2 \: m}^{n_1'+n_2'-k_1-k_2 \: m'} \; .
\end{eqnarray}
\end{widetext}
Here, as well as above in Appendix~\ref{Ap:Tr},  
$V_{n \, m}^{n'\, m'} = \delta_{n+m,n'+m'} F_{n \, m}^{n'\, m'}$
is the single-particle matrix element defined, e.g., in 
Eqs.~(43) and (44) of Ref.~\onlinecite{Dzy00}.
Two different expressions (\ref{matApT}) and (\ref{matAp1}) 
for the same matrix elements
are connected with two different representations of the $SU(2)$ group. 
Their equivalence at low finite orders can be checked directly, term by term, 
and, generally, by the methods indicated in Ref.~\onlinecite{Gil05}.
The form (\ref{matAp1}) appears to be more suitable for numerical calculations.

%\mbox{}
%\vspace{50pt}
%\mbox{}

%%%%%%%%%%%%%%%%%%%%%%%%%%%%%%%%%%%%%%%%%%%%%%%%%%%%%%%%%%%%%%%%%%%%%%%%%%%%

%*************************************************************

\begin{thebibliography}{99}
\bibitem{Note:Charge}For positive charges $q_1>0$ and $q_2>0$, the  
                     eigenenergies are the same as given in the present paper, 
                     while the wavefunctions can be 
                     obtained by complex conjugation.
\end{thebibliography}
\end{document}